\documentclass{elsart}

\usepackage{amssymb}
\usepackage{graphicx}
\usepackage{dcolumn}
\usepackage{bm}

\begin{document}

\newcommand{\FTxSx}{FeSe$_{1-x}$Te$_x$}
\newcommand{\FTS}{FeSe$_{0.44}$Te$_{0.56}$}

\begin{frontmatter}



\title{The crystal structure of \FTS}


\author{M.~Tegel,}
\author{C.~L\"ohnert,}
\author{D.~Johrendt\corauthref{cor1}}
\ead{johrendt@lmu.de}
\corauth[cor1]{Corresponding author}

\address{Department Chemie, Ludwig-Maximilians-Universit\"{a}t M\"{u}nchen,\\ Butenandtstr. 5-13 (Haus D), 81377 M\"{u}nchen, Germany}

\begin{abstract}

The crystal structure of the superconductor \FTS~was redetermined by high-resolution X-ray single crystal diffraction at 173~K (anti-PbO-type, $P4/nmm$, $a$ = 3.7996(2), $c$ = 5.9895(6)~{\AA}, $R1$ = 0.022, $wR2$ = 0.041, 173~$F^2$). Significantly different $z$-coordinates of tellurium and selenium at the $2c$ site are clearly discernible and were refined to $z_{\rm Te}$=0.2868(3) and $z_{\rm Se}$=0.2468(7). Thus the chalcogen heights differ by 0.24 {\AA} and the Fe--Se bonds are by 0.154~{\AA} shorter than the Fe--Te bonds, while three independent (Te,Se)--Fe--(Te,Se) bond angles occur. An elevated $U_{33}$ displacement parameter of the iron atom is suggestive of a slightly puckered Fe layer resulting from different combinations of Se or Te neighbors. Such strong disorder underlines the robustness of superconductivity against structural randomness and has not yet been considered in theoretical studies of this system.

\end{abstract}

\end{frontmatter}


\section{Introduction}

$\beta$-FeSe with the tetragonal anti-PbO-type structure \cite{Haegg-1933,Okamoto-1991} can be considered as the archetypal iron-based superconductor. Consisting solely of superconducting layers without separating atoms or building blocks, this seems to be the ideal system to study the underlying physics, which turned out to have many   analogies to the iron arsenide superconductors \cite{Hosono-2008,Rotter-2008-1,Wang-2008}. However, soon after the discovery of superconductivity in FeSe~\cite{Hsu-2008} with $T_c$ = 8~K, a set of studies revealed differences between iron selenide and pnictide materials. Examples are the extreme sensitivity of superconductivity to the stoichiometry \cite{Cava-2009-1,Cava-2009-2}, the huge increase of $T_c$ under pressure \cite{Mizuguchi-2008,Cava-2009-3}, the absence of long-range magnetic ordering and the orthorhombic or even monoclinic symmetry of the superconducting phase \cite{Cava-2009-4}. In contrast to the selenide, the binary iron telluride with anti-PbO-type structure \cite{Leciejewicz-1963} is not superconducting even under pressure \cite{Mizuguchi-2009}. This may be due to the fact that stoichiometric FeTe does not exist, but only Fe$_{1+\delta}$Te with $\delta~\approx$~0.05-0.15. The excess iron atoms between the layers may be detrimental to superconductivity \cite{Han-2009}. On the other hand, superconductivity survives in the solid solution \FTxSx~with a maximum $T_c$ of about 14 K close to FeSe$_{0.5}$Te$_{0.5}$ \cite{Yeh-2008}.

The \FTxSx~system has been intensively studied with respect to the interplay between structural or magnetic degrees of freedom and superconductivity. It is accepted that magnetic fluctuations play an important role, and calculations revealed a strong sensitivity of the magnetic moment on the so-called ``chalcogen height'', $i.~e.$ the distance of the Se/Te atoms from the plane of iron atoms \cite{Moon-2009}. According to this, the magnetic exchange parameter $J$ changes by a factor of 2-5 when the height varies by only 0.1~{\AA}. Such subtle dependencies of the magnetic and superconducting properties are also known from the iron pnictides \cite{Lebegue-2009}, which re-emphasizes the importance of accurate and correct structural data. However, against the background of several published crystal structures of \FTxSx \cite{Rosseinsky-2009}, it is surprising that even simple crystal chemical aspects have been completely disregarded. Selenium and tellurium have quite different ionic radii: $r_{\rm{Se}^{2-}}$~=~1.98~{\AA} and $r_{\rm{Te}^{2-}}$~=~2.21~{\AA} \cite{Shannon}. Therefore, one cannot expect that both ions occupy the $2c$ position with the same $z$-coordinate in the layered PbO-type structure and we wondered why the structure is described with Se and Te in the same position in many publications. Only in a recent preprint by Lehman $et~al.$ \cite{Lehman-2009} were different $z$-coordinates of FeSe$_{0.5}$Te$_{0.5}$ determined laboriously and with low precision from an analysis of the pair density function (PDF) obtained from neutron powder diffraction data.

In this paper we present a single-crystal X-ray diffraction study of \FTS, which easily reveals the distinct $z$-coordinates of Se and Te with a one order of magnitude better accuracy compared to the PDF method. We identify a quite large degree of structural disorder, which has been neglected in recent studies and emphasize the robustness of superconductivity against randomness.

\section{Experimental details}

\begin{figure}[h]
\center{
\includegraphics[width=90mm]{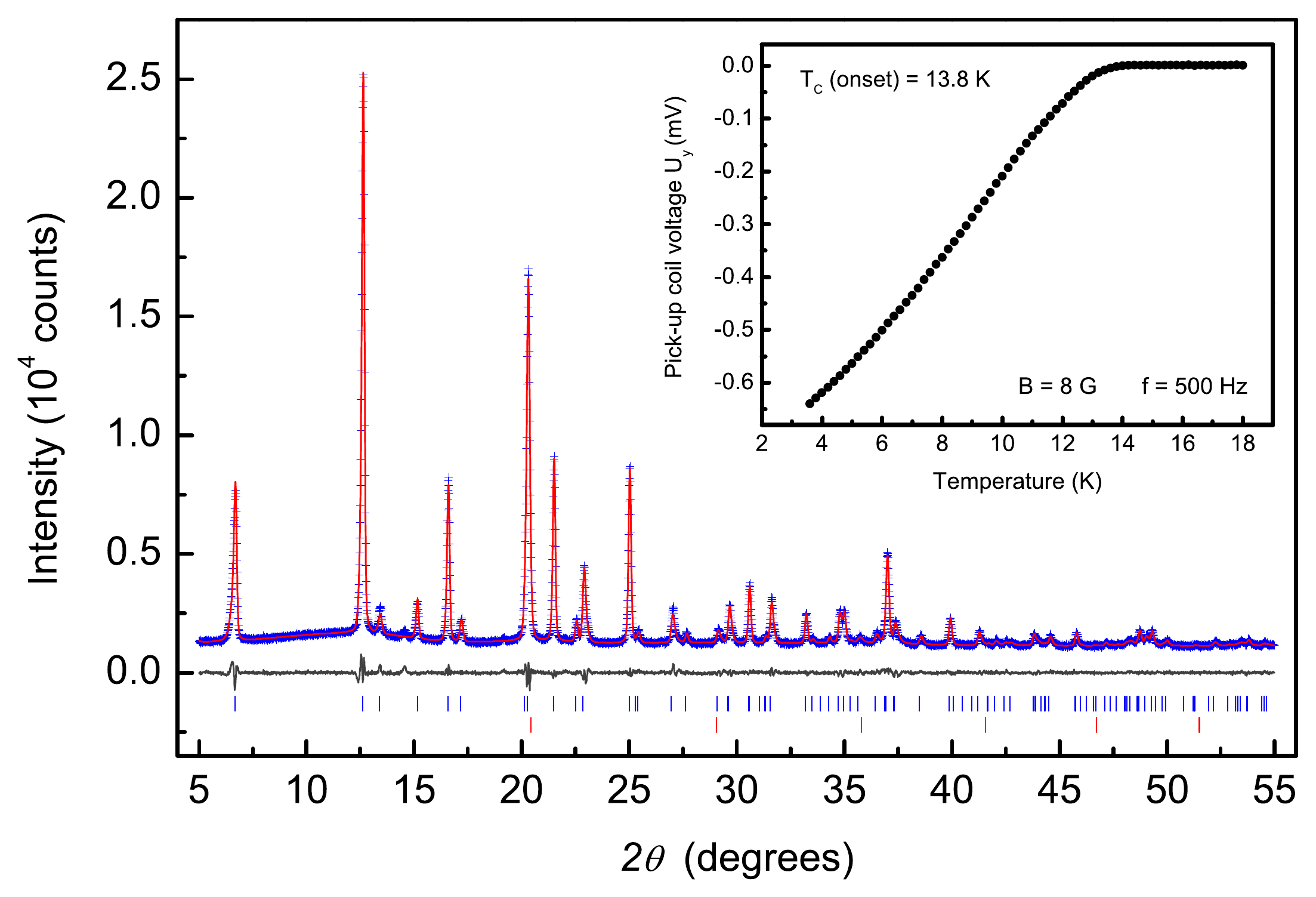}
\caption{X-ray powder pattern (blue) and Rietveld fit (red) of FeSe$_{0.4}$Te$_{0.6}$. Inset: AC susceptibilty measurement showing the onset of superconductivity at $\approx$~14 K}
\label{fig:measurement}
}
\end{figure}

The sample with the nominal composition FeSe$_{0.4}$Te$_{0.6}$ was synthesized by heating the elements (purity $>$ 99.8\%) at 973~K for 40~h in an alumina crucible, sealed in a silica tube under an atmosphere of purified argon. After cautious homogenization using an agate mortar in an argon-filled glove box, the sample was heated again at 1012~K for 30~h. The X-ray powder diffractogram (STOE Stadi-P, Mo-K$_{\alpha_1}$ radiation) showed the expected pattern of \FTxSx. A Rietveld refinement using the TOPAS package \cite{Topas} (see Figure~\ref{fig:measurement}) revealed only traces of impurity phases. The refined lattice parameters are $a = 3.8061(2)$, $c = 6.0871(3)$~{\AA}. Superconductivity was confirmed by measuring the AC susceptibility (inset in Figure~\ref{fig:measurement}). A relatively broad transition with an onset of $T_c$ at 14~K was found. Such broad transitions in \FTxSx~superconductors have also been observed by other groups \cite{Fang-2008,Rosseinsky-2009}, and may be due to a certain sample inhomogeneity. A small plate-like single crystal of $\approx 15 \times 20 \times 30$ microns was selected from the polycrystalline sample and checked by Laue photographs using white radiation from a Mo anode. Diffraction intensity data up to $2\theta$ = 90$^{\circ}$ (0.5~{\AA} resolution) were collected at 173~K on an Oxford Xcalibur 4-circle diffractometer equipped  with a CCD detector. Graphite-monochromized Mo-K$\alpha$ radiation from a conventional sealed tube was used. The measured intensities were carefully corrected for absorption effects. The atom positions from \cite{Rosseinsky-2009} were used as starting parameters ($z_{\rm Se/Te}$ = 0.27388) and refined by the least-squares method using the JANA2006 program package \cite{Jana}. The positional and thermal parameters of Se and Te were refined independently, while their occupation parameters were constrained to unity.

\section{Structure refinement}

The initial refinement of the data using one $z$-coordinate already resulted in small residuals of $R1$ = 0.029 and $wR2$ = 0.051. But the inspection of the anisotropic displacement parameters showed a three times larger displacement parameter U$_{33}$ when compared with U$_{11}$. It is known that such elongations can be artifacts from insufficient absorption corrections. For this reason we performed several refinements using different absorption models. This revealed only a small dependency, which cannot explain a factor of three in the anisotropy. In the next step we enabled the independent refinement of the Se and Te $z$-coordinates, which immediately converged to different values for Se ($\approx$~0.246) and Te ($\approx$~0.286), while the residuals dropped to $R1$ = 0.023 and $wR2$ = 0.05. Moreover, the anisotropy of the still combined (Se,Te) thermal ellipsoid was reduced from $\approx$~3 to $\approx$~1.3. The necessary independent refinement of the anisotropic displacement is difficult, because the $z$-coordinates and the U$_{33}$ parameters are strongly correlated. However, due to the high-resolution data (2$\theta_{max} \approx$ 90$^{\circ}$), the correlations remained acceptable and did not destabilize the refinement. Final cycles converged to residuals $R1$ = 0.022, $wR2$ = 0.041 and GooF = 0.98. A careful check of the $\Delta F$ Fourier map revealed no residual density from additional Fe atoms between the layers. Regarding the Fe atom in the layer, all refinements showed an elongation of the thermal ellipsoid along $z$ with an anisotropy of $\approx$~2. We suggest that this is the response of the iron atom to the different (Se,Te) coordinations, tantamount to a small puckering of the Fe plane. A summary of the crystallographic data is compiled in Table~\ref{tab:Crystallographic}, the final atom position parameters and anisotropic displacements are given in Table~\ref{tab:Positions} and relevant bond distances and angles are collected in Table~\ref{tab:Distances}.

Further details of the crystal structure investigation in CIF format may be obtained from Fachinformationszentrum Karlsruhe, 76344 Eggenstein-Leopoldshafen, Germany (fax: (+49)7247-808-666; e-mail: crysdata@fiz-karlsruhe.de, http://www.fiz-karlsruhe.de/request\_for\_deposited\_data.html) on quoting the CSD number 421334.

\begin{figure}[h]
\center{
\includegraphics[height=80mm]{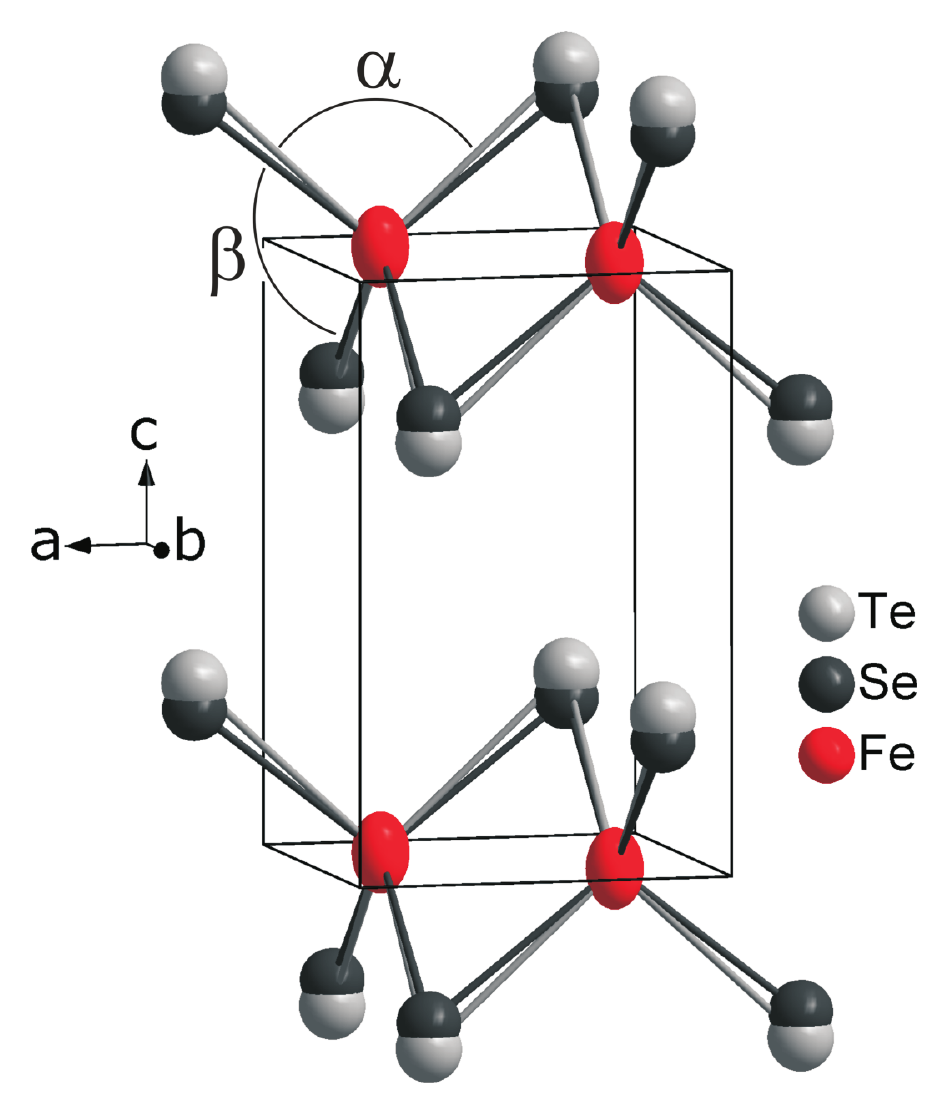}
\caption{Crystal structure of \FTS~with Se/Te split positions. (space group $P4/nmm$, thermal ellipsoids of 95\% probability)}
\label{fig:Structure}
}
\end{figure}

\section{Discussion}

The crystal structure of \FTS~is depicted in Figure~\ref{fig:Structure}. The redetermination clearly revealed the distinct positions of Se and Te as expected from crystal chemical reasons and agrees with the assumption made in Ref.~\cite{Lehman-2009}. The resulting ``chalcogen heights'' are $h_{\rm Se}$ = 1.478~{\AA} and $h_{\rm Te}$ = 1.718~{\AA}. In terms of bond lengths, the Fe--Se bond length is by 0.154~{\AA} shorter than the Fe--Te bond. In comparison to the binary compounds, the Fe--Se bond is slightly longer (by 1.6~\%) than in FeSe \cite{Haegg-1933} and the Fe--Te bond is slightly shorter (by 1.8~\%) than in Fe$_{1+\delta}$Te \cite{Leciejewicz-1963}. This is plausible from crystal chemistry, in contrast to a mean Fe--(Se,Te) bond length if only one $z$-parameter is used. Also the (Se,Te)--Fe--(Se,Te) bond angles depend significantly on the $z$-coordinates. Iron occupies a position with $\overline{4}m2$ symmetry; therefore the angle which is bisected by the $c$-axis ($\alpha$) and the angle which is bisected by the $ab$-plane ($\beta$) are related by $\cos(\beta) = -\frac{1}{2}[1- \cos(\alpha)]$. In the present case, three independent angles $\alpha$ have to be considered, namely Te--Fe--Te, Se--Fe--Se and Te--Fe--Se. We have also listed the dependent angles $\beta$ in Table~\ref{tab:Distances} for the sake of completeness.

In the case of the iron pnictide superconductors, it has been argued that the geometry of the Fe$Pn_4$ tetrahedra plays an important role. $T_c$ is seemingly maximized when the bond angles are close to the ideal tetrahedral angle of 109.47$^{\circ}$ \cite{Lee-2008}. The latter has not been observed in the Fe$Ch$ systems, but the large enhancement of $T_c$ under pressure suggests again that details of the structure are involved in the mechanism of superconductivity. Furthermore, it is undoubted that the coordination of the iron atoms controls the Fermi surface topology, which is very similar to the Fe$Pn$ superconductors \cite{Subedi-2008}. Also the magnetic moment depends strongly on the height of the pnictogen \cite{Yildirim-2009} and chalcogen \cite{Moon-2009} atoms; thus the difference $\Delta h_{\rm Te-Se}$ = 0.24~{\AA} as a consequence of the split positions is a strong effect. With respect to the studies on structural effects in the \FTxSx~system reported so far \cite{Rosseinsky-2009,Moon-2009,Bao-2009}, we want to point out that the changes in the geometry by the split position of Se and Te are at least one magnitude larger than the effects induced by temperature or pressure.

A further implication of our results concerns the discussion about the robustness of superconductivity in iron-based materials in the framework of the s$_{\pm}$-scenario \cite{Mazin-2008}. The latter is expected to be rather fragile against impurities or other kinds of randomness \cite{Kontani-2009}. In this context, it is remarkable that $T_c$ increases from 8~K in FeSe to 14~K in \FTxSx~$(x\approx 0.5)$ notwithstanding the strong geometrical disorder caused by the Te-doping at the very different $z$-coordinates. Detailed calculations taking this into account are necessary to fully understand the consequences of this disorder on the superconducting properties of \FTxSx.

\section{Summary}

We have redetermined the crystal structure of \FTS~by refinements of high-resolution single-crystal X-ray data and found distinctively different $z$-coordinates for Se and Te. This leads to a lower local symmetry of the iron atoms by splitting of the Fe--$Ch$ bond lengths by 0.154~{\AA} and to different chalcogen heights of 1.478~{\AA} and 1.718~{\AA}. Such large effects have not yet been considered in previous calculations of electronic and magnetic properties, which are known to be very sensitive to bond lengths and angles. Since the value of $T_c$ in \FTxSx~increases up to $x \approx 0.5$ despite the significant Se/Te disorder, our results point out the robustness of superconductivity in iron-based materials against structural randomness.

\bibliographystyle{elsart-num}


\begin{table}[h]
\caption{Crystallographic data$^a$}
\label{tab:Crystallographic}
\begin{tabular}{ll}
\hline
Empirical formula         & \FTS\\
Molar mass (g/mol)        & 162.2\\
Crystal system, space group               & Tetragonal, $P4/nmm$, No.~129\\
\textit{a,~c} ({\AA})     & 3.7996(2),~5.9895(6)\\
Cell volume ({\AA}$^3$)   & 86.47(1)\\
Calc. density (g/cm$^3$),~\textit{Z}              & 6.23,~2\\
Radiation type, $\lambda$~({\AA})                 & Mo-K$_{\alpha}$,~0.7107\\
Absorption coefficient (mm$^{-1}$)                & 26.6\\
Crystal size ($\mu m^3$)                          & 15~$\times$~30~$\times$~20\\
Absorption correction                             & multi-scan\\
Transmission (min,~max)                           & 0.39,~1.00 \\
2$\theta$-range                                          & 10.7 - 90.23\\
Total no. Reflections                             & 913 \\
Independent Reflections,~R$_{\rm int}$            & 246,~0.037 \\
Reflections with $I>3\sigma(I)$,~R$_{\sigma}$     & 173,~0.038 \\
Refined parameters, GooF on $F^2$                 & 9,~0.98 \\
$R1,~wR2$~~($I>3\sigma(I)$)                      & 0.022,~0.041\\
$R1,~wR2$~~(all data)                            & 0.0417,~0.043\\
Largest residual peak,~hole~(e{\AA}$^3$)          & 0.98,~$-$1.59\\
\hline
\multicolumn{2}{l}{$^a$Numbers in parentheses are standard deviations of the last significant digits}
\end{tabular}
\end{table}

\begin{table}[h]
\caption{Atom positions and displacement parameters ({\AA}$^2$)$^b$}
\label{tab:Positions}
\begin{tabular}{llllllll}
\hline
Atom & Wyck. & occ. & $x$ & $y$ & $z$ & U$_{11}$ & U$_{33}$ \\
\hline
Fe   & 2$a$ & 1 & $\frac{3}{4}$ & $\frac{1}{4}$ & 0 &  0.0064(2) & 0.0132(4)\\
Se   & 2$c$ & 0.44(1) & $\frac{1}{4}$ & $\frac{1}{4}$ & 0.2468(7) & 0.0080(6) & 0.0116(7)\\
Te   & 2$c$ & 0.56(1) & $\frac{1}{4}$ & $\frac{1}{4}$ & 0.2868(3) & 0.0084(5) & 0.0118(7)\\
\hline
\multicolumn{6}{l}{$^{b}$U$_{22}$ = U$_{11}$; U$_{12}$ = U$_{13}$ = U$_{23}$ = 0}\\
\end{tabular}
\end{table}

\begin{table}[h]
\caption{Atomic distances ({\AA}) and bond angles (deg, first row: $\alpha$, second row: $\beta$)$^c$}
\label{tab:Distances}
\begin{tabular}{lllll}
\hline
Fe--Se  & 2.407(3)   & Se--Fe--Se 104.1(1) & Te--Fe--Te 95.75(4)  & Se--Fe--Te 99.99(9)\\
Fe--Te  & 2.561(1)   & Se--Fe--Se 112.15(6) & Te--Fe--Te 116.74(2) & Se--Fe--Te 114.32(5)\\
\hline
\multicolumn{5}{l}{$^c\cos(\beta) = -\frac{1}{2}[1- \cos(\alpha)$]}
\end{tabular}
\end{table}

\end{document}